\documentstyle[epsfig,psfig]{mn}

\input{epsf}

\begin{document}

%TITLES AND AUTHORS
\title
[A measurement of $\Lambda$ from the QSO power spectrum $P^S(k_{\parallel},\mathbf{k}_{\perp})$]
{
The 2dF QSO Redshift Survey - XIII. A measurement of $\Lambda$ from the
QSO power spectrum, $P^S(k_{\parallel},\mathbf{k}_{\perp})$
}

\author[P.~J. Outram et~al. ]
{
P.~J. Outram$^1$,  T. Shanks$^1$,  B.~J. Boyle$^{2}$, S. M. Croom$^{2}$, Fiona Hoyle$^{3}$, 
\newauthor   N.~S. Loaring$^{4}$, L. Miller$^{5}$, \& R.~J. Smith$^{6}$
\\
1 Department of Physics, Science Laboratories, South Road, Durham, DH1 3LE, U.K.
\\
2 Anglo-Australian Observatory, PO Box 296, Epping, NSW 2121, Australia\\
3 Department of Physics, Drexel University, 3141 Chestnut Street, Philadelphia, PA 19104, U.S.A.\\
4 Mullard Space Science Laboratory, University College London, Holmbury St. Mary, Dorking, Surrey,
RH5 6NT, U.K.\\
5 Department of Physics, Oxford University, Keble Road, Oxford, OX1 3RH, U.K.\\
6 Liverpool John Moores University, Twelve Quays House, Egerton Wharf, Birkenhead, CH41 1LD, U.K. \\
}

\maketitle 
 
\begin{abstract}

We report on measurements of the cosmological constant,
$\Lambda$, and the redshift space distortion parameter
$\beta=\Omega_m^{0.6}/b$, based on an analysis of the QSO power
spectrum parallel and perpendicular to the observer's line of sight, 
$P^S(k_{\parallel},\mathbf{k}_{\perp})$, from the final catalogue of
the 2dF QSO Redshift Survey.
We derive a joint $\Lambda - \beta$ constraint  from the geometric and
redshift-space distortions in the power spectrum. By combining this
result with a second constraint based on mass clustering evolution, we
break this degeneracy and obtain strong constraints on both parameters.
Assuming a flat ($\Omega_{\rm m}+\Omega_{\Lambda}=1$) cosmology and a
$\Lambda$ cosmology $r(z)$ function to convert from redshift into comoving
distance, we find
best fit values of 
$\Omega_{\Lambda}=0.71^{+0.09}_{-0.17}$ and
$\beta_q(z\sim1.4)=0.45^{+0.09}_{-0.11}$. Assuming instead an EdS cosmology
$r(z)$ we find that the best fit model obtained, with $\Omega_{\Lambda}=0.64^{+0.11}_{-0.16}$ and
$\beta_q(z\sim1.4)=0.40^{+0.09}_{-0.09}$, is consistent with the $\Lambda$ $r(z)$
results, and inconsistent with a
$\Omega_{\Lambda}=0$ flat cosmology at over 95 per cent confidence. 

\end{abstract}


\section{Introduction}
Observations of high redshift Type Ia Supernovae (SNIa) have recently
generated much excitement and interest in $\Lambda$CDM cosmological
models, deriving strong constraints on the cosmological constant,
$\Lambda$ (e.g. Perlmutter et al. 1999; Riess et al. 1998). There are
still uncertainties regarding the standard candle assumption that
underlies this approach, however, and we don't yet have a good
understanding of SNIa physical processes. It is therefore
important to develop independent ways to constrain $\Lambda$ in order
to confirm this tantalizing result, as different methods suffer from different systematic uncertainties.

Different astronomical datasets are sensitive to different combinations of the
cosmological parameters. For example, the recent WMAP cosmic microwave
background (CMB) observations
provide a strong constraint on $\Omega_mh^2$ (Spergel et al. 2003),
whereas the shape of the matter power spectrum, probed by either
galaxies (Percival et al. 2001) or QSOs (Outram et al. 2003), is sensitive to $\Omega_mh$. By
combining several approaches we can break these degeneracies and hence
derive much stronger constraints on the model that best describes our
Universe (e.g. Efstathiou et al. 2002). Neither the CMB
nor the large-scale structure observations, however, are directly
sensitive to $\Omega_{\Lambda}$, and can only indirectly infer its
value when the two datasets are combined.

Alcock \& Paczy\`{n}ski (1979) suggested that $\Lambda$ might be
measured directly from redshift-space distortions in the shape of
large-scale structure, by making the simple assumption that clustering
in real-space is on average spherically symmetric. Geometric
distortions occur if the wrong cosmology is assumed, due to the
different dependence on cosmology of the redshift-distance relation
along and across the line of sight, and from the size of these
geometric distortions, the true cosmology can be determined.

One approach to implementing this test is to use the distribution of
 QSOs to probe large-scale structure. By comparing the clustering of QSOs along and across the line of sight and
modelling the effects of peculiar velocities and bulk motions in
redshift space (Ballinger et al. 1996, Matsubara \& Suto 1996),
geometric distortions can be detected. 

The ideal sample for this is the recently completed 2dF QSO Redshift
Survey (2QZ). Data for 2QZ were obtained using the AAT 2dF facility. The
completed survey comprises some 23000
 $b_{\rm J}<20.85$ QSOs in two $5\times75\deg^2$
declination strips, one at the South Galactic Pole and one in an
equatorial region in the North Galactic Cap spanning the redshift range
$0.3\la z\la 2.5$. The 2QZ catalogue and spectra were released in July
2003 (Croom et al. 2003) and can be obtained at {\tt
  http://www.2dfquasar.org}.

Outram et al. (2001) produced a preliminary analysis of
redshift-space distortions in the power spectrum,
$P^S(k_{\parallel},\mathbf{k}_{\perp})$, of the incomplete 10k 2QZ
survey (Croom et al. 2001). 
To help develop the method, and to test estimators designed to discriminate the effects of
infall (measured via the parameter $\beta \sim \Omega_m^{0.6}/b$),
small-scale velocity dispersion, and geometric distortions, a
simulation of the 2QZ was used. Three light
cone strips of the redshift survey were simulated, using the Virgo
Consortium's huge Hubble Volume N-body $\Lambda$CDM simulation
(Frenk et al. 2000, Evrard et al. 2002). The
effects of infall and geometry on the clustering distribution are very
similar, and this approach alone could therefore only provide a degenerate
constraint on $\Lambda$ and $\beta$. Outram et al.  found, however, that this
degeneracy could be broken by jointly considering a second constraint
based on mass clustering evolution. 
Using the Hubble Volume simulation, Outram et al.  predicted that this method should
constrain $\beta$ to approximately $\pm0.1$, and $\Omega_{\Lambda}$ to
$\pm0.25$ using the final 2QZ catalogue.
A complementary approach, using the two-point correlation function,
$\xi(\sigma,\pi)$, to investigate redshift-space distortions in QSO
clustering can also be considered (Hoyle et al. 2002). Here we shall apply the 
analysis developed in Outram et al. to the final 2QZ sample, and hence
derive this significant new constraint on $\Lambda$.

\section{Power Spectrum Analysis}\label{secpk}

\begin{figure}
\vspace{-0.5cm}
\centerline{\hbox{\psfig{figure=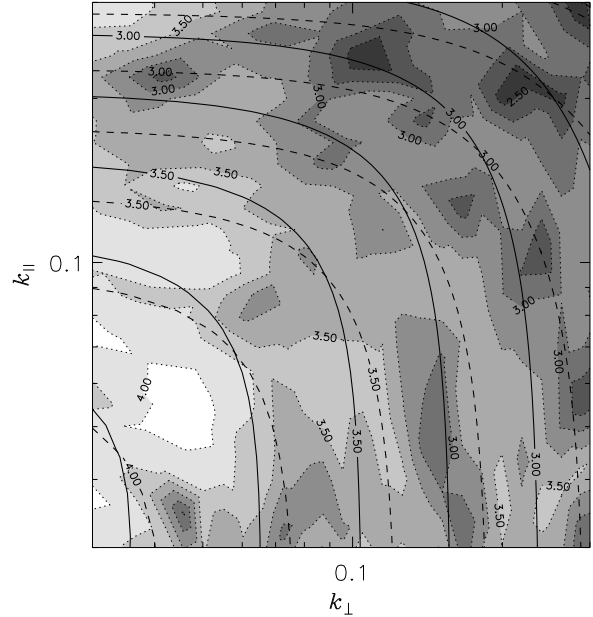,width=9cm}}}
\vspace{-0.3cm}
\caption{
  $P^S(k_{\parallel},\mathbf{k}_{\perp})$ determined from the 2QZ
  Catalogue assuming an EdS  $r(z)$. Filled contours of constant ${\mathrm log}
  (P(k)/h^{-3}$Mpc$^3)$ are shown as a function of
  $k_{\parallel}/h$\,Mpc$^{-1}$ and ${\mathbf
    k}_{\perp}/h$\,Mpc$^{-1}$. Overlaid are the best fit model (solid
  contours) with $\beta = 0.40$ and $\Omega_{\Lambda} = 0.64$,  and for
  comparison, the best fitting flat $\Omega_{\Lambda} = 0$ model with
  $\beta = 0.27$ (dashed contours).
}
\label{figdataom}
\end{figure}

We reapply the analysis described in Outram et al. (2001) to the final
2QZ catalogue, containing 22652 QSOs (only those QSOs with quality 1 are
used in this analysis; see Croom et al. (2003) for further details). For
convenience we review the main details below.

\subsection{The 2QZ selection function}

We account for the various selection effects introduced into the 2QZ
catalogue by generating a catalogue of random points
that mimics the angular and radial selection functions of the QSOs but
otherwise is unclustered. Details of the QSO selection function used
can be found in Section 2.1 of Outram et al. (2003). We are measuring QSO clustering 
as a function of comoving distance, and
so need to assume a cosmology in order to convert from redshift into comoving
distance.  Geometric
distortions in the shape of
large-scale structure  occur if this cosmology is wrong, due to the
different dependence on cosmology of the redshift-distance relation
along and across the line of sight. From the size of these
geometric distortions, measured in the QSO power spectrum,
$P^S(k_{\parallel},\mathbf{k}_{\perp})$, the true cosmology can be
determined. The choice of adopted cosmology shouldn't matter in this
analysis; adopting an incorrect cosmology does not
affect the validity of this result.
To check for consistency, however, we consider two possibilities; an Einstein-de Sitter ($\Omega_{\rm
  m}$=1.0, $\Omega_{\Lambda}$=0.0) cosmology $r(z)$ (EdS hereafter) and
an $\Omega_{\rm m}$=0.3, $\Omega_{\Lambda}$=0.7 cosmology $r(z)$
($\Lambda$ hereafter). To limit incompleteness, we restrict our
analysis to $0.3<z<2.2$. This restricts our sample to 19549 QSOs with a
mean redshift of $\bar z\sim1.4$.

\subsection{Power spectrum estimation}\label{pse}

\begin{figure}
\vspace{-0.5cm}
\centerline{\hbox{\psfig{figure=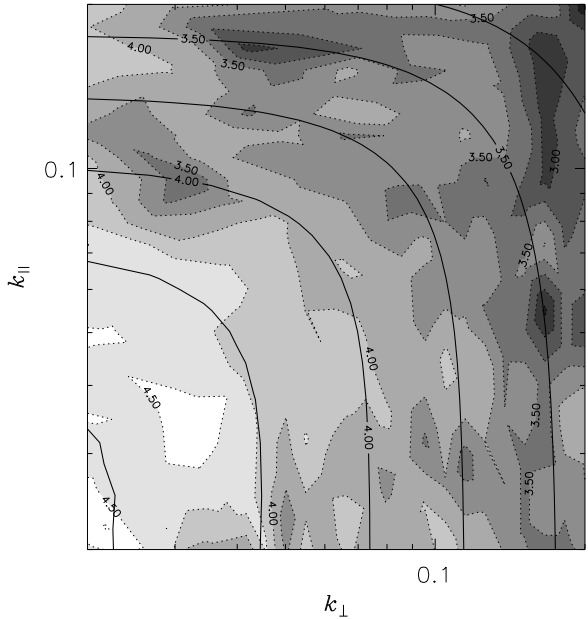,width=9cm}}}
\vspace{-0.3cm}
\caption{
  $P^S(k_{\parallel},\mathbf{k}_{\perp})$ determined from the 2QZ
  Catalogue assuming a $\Lambda$ $r(z)$. Filled contours of constant ${\mathrm log}
  (P(k)/h^{-3}$Mpc$^3)$ are shown as a function of
  $k_{\parallel}/h$\,Mpc$^{-1}$ and ${\mathbf
    k}_{\perp}/h$\,Mpc$^{-1}$. Overlaid is the best fit model with $\beta = 0.45$ and $\Omega_{\Lambda} = 0.71$.
}
\label{figdatalam}

\end{figure}

The power spectrum estimation is carried out as described in Outram,
Hoyle \& Shanks (2001). 
When calculating $P^S(k_{\parallel},\mathbf{k}_{\perp})$, information about the line of sight must be retained. To achieve this the data are divided into subsamples that subtend a small solid angle on the sky, and the distant-observer approximation is then applied. The data from each declination strip are split into 8 regions each approximately $5^{\circ} \times 10^{\circ}$ during this analysis. 
Each region is embedded
 into a larger cubical volume, which is rotated  such that the central
 line of sight lies along the same axis of the cube in each case. The
 density field is binned onto a 256$^3$ mesh, using nearest grid-point
 assignment.  The power spectrum of each region is estimated using a
 Fast Fourier Transform (FFT), and the average of the resulting power
 spectra is taken. The results, binned logarithmically into
 $k_{\parallel}$ and $\mathbf{k}_{\perp}$, are plotted in
 Figure~\ref{figdataom} (assuming an EdS $r(z)$) and   Figure~\ref{figdatalam}
 (assuming a $\Lambda$ $r(z)$). 

\subsection{The window function}
\begin{figure}
\vspace{-0.7cm}
\centerline{\hbox{\psfig{figure=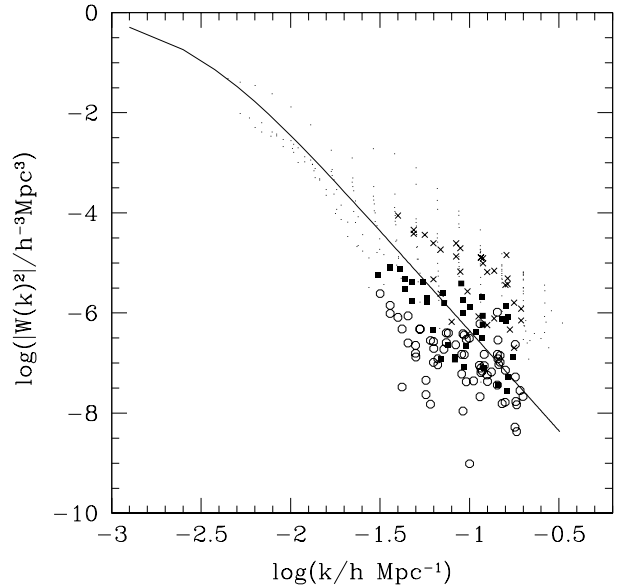,width=9.5cm}}}
\vspace{-0.7cm}
\caption{The  power spectrum of the window function of the 2QZ
  catalogue, assuming $\Lambda$ $r(z)$, shown as a function of $k$. The
  solid squares show the points which were used in the analysis, with
  wavenumbers with $k_{\parallel}>0.02, |{\mathbf k}_{\perp}|>0.02$,
  and $k<0.2$. The dots show the power spectrum at smaller wavenumbers;
  excluded from this analysis.
The window function  is a steep power law, varying as $\sim k^{-3}$.
For comparison, the open circles show the power spectrum of the window
function from the analysis of the 2QZ 25k  mock catalogue, and the
crosses  show the power spectrum of the window
function from the analysis of the 2QZ 10k catalogue (Outram et
al. 2001). The points are significantly lower than those of the 10k
catalouge, due to the increased coverage, but still slightly higher than those of the mock
catalogue due to the varying completeness pattern that still affects
the finished survey (Croom et al. 2003).  The overlaid line is a fit to
the window function from the usual power spectrum analysis, binning in
shells of $k$ (Outram et al. 2003).
}
\label{figwin1d}
\end{figure}

The measured power spectrum is convolved with the power spectrum
of the window function. The power spectrum of the window function is shown in
figure~\ref{figwin1d}. We compare the 2QZ window function to that of
the  10k catalogue and the 25k  mock catalogue determined in Outram et
al. (2001). Due to the varying observational and
spectroscopic incompleteness imprinted on the angular distribution of
the 2QZ QSOs, the points lie slightly above those of the mock catalogue
(where it was assumed that the angular distribution of the final
catalogue would be uniform), but significantly lower than those of the
10k catalogue, due to the increased coverage.
Outram et al. (2001; Section~2.3) adopted limits of $k_{\parallel}$ \&
$|\mathbf{k}_{\perp}|$ $ \ge0.02\;h\;$Mpc$^{-1}$ for the analysis of
$P^S(k_{\parallel},\mathbf{k}_{\perp})$ to avoid any problems
due to the effect of the window function on large scales. Whilst
slightly higher than that of the mock catalogue, the window function is
still sharply peaked in ${\mathbf k}$-space, and so we adopt the same limits in
the case of the $\Lambda$ $r(z)$ analysis. Any window function effects should still be negligible.
 Assuming an EdS $r(z)$,
however, the comoving distances probed are smaller, and hence the
window function affects smaller scales. The appropriate limits to adopt
in this case are $k_{\parallel}$ \&
$|\mathbf{k}_{\perp}|$ $ \ge0.03\;h\;$Mpc$^{-1}$.

\subsection{Error estimates}\label{errsec}

A different error estimate is adopted in this paper from that of Outram
et al. (2001). They adopted two methods to estimate the error in the
power spectrum measurements; either via the dispersion between the
power spectrum measurements from 
three declination strips of the Hubble Volume simulation, or the error
estimate obtained using the method of Feldman, Kaiser \& Peacock (FKP;
1994).
The two estimates agree fairly well, but with large scatter. The FKP
error is, on average, slightly lower than the error estimated from the
dispersion betwen the three mock strips.  Neither approach is ideal; the
error estimated from the dispersion has a high uncertainty due to the
small number of mock catalogues, and relies on the power of the mock catalogues
being similar to that of the 2QZ catalogue,  whereas the FKP error estimate assumes a thin shell geometry, and can under-estimate the error on some scales, due, for example, to the onset of non-linearity.

In order to apply the
 distant-observer approximation (see Section~\ref{pse}) the QSO power
 spectrum was estimated separately in  sixteen individual $5^{\circ}
 \times 10^{\circ}$ regions. For this paper we have chosen to estimate the errors self-consistently
from the data by considering the dispersion in these sixteen power
spectrum measurements.  As these regions are taken from just two
contiguous strips, they lie next to each other, and
are not entirely independent. It is therefore possible that they
slightly underestimate the error in the power estimate, especially on
large scales. Whilst we believe this effect would be small on the scales
of interest, and hence do not consider it further in this paper, to
test this fully would require several independent simulations of the
survey, which are not currently available due to the large volume probed.
This error estimate agrees well, on
 average, with estimates obtained from the Hubble Volume simulation (see section~2.4
 of Outram et al. 2001), but is slightly larger than the FKP error estimate.

\section{Modelling Redshift-space Distortions}\label{secmod}
The power spectrum model incorporating redshift distortions that we fit
to the data is presented in Ballinger, Peacock and Heavens (1996), and
briefly summarized below.

The power spectrum analysis was carried out assuming either an $\Omega_{\rm
  m}$=1.0, $\Omega_{\Lambda}$=0.0 cosmology  or
an $\Omega_{\rm m}$=0.3, $\Omega_{\Lambda}$=0.7 cosmology. If the true
cosmology differs from this then geometric distortions will be
introduced into the clustering pattern, due to the different dependence on cosmology of the redshift-distance relation along and across the line of sight. Our distance calculations will be wrong by a factor $f_{\perp}$ perpendicular to the line of sight, and $f_{\parallel}$ along the line of sight (as defined in Ballinger et al. 1996). Thus we can define a geometric flattening factor:

\begin{equation}\label{flateqn}
F(\Omega_{\Lambda},z)=\frac{f_{\parallel}}{f_{\perp}}%=1+\frac{3}{4}\Omega_{\Lambda}z + O(z^2)
\end{equation}
(Ballinger et al. 1996). The effect this has on the power spectrum is given by
\begin{equation}\label{geomeqn}
P_{\mathrm anisotropic}(k_{\parallel},{\mathbf k}_{\perp})= 
\frac{P_{\mathrm true}(k)}{f_{\perp}^{3+n}F}  
\left[1+\mu^2\left(\frac{1}{F^2}-1\right)\right]^{\frac{n}{2}}
\end{equation}
(Ballinger et al. 1996) where $\mu=k_{\parallel}/k$, and $n$ is the
spectral index of the power spectrum.  By measuring the size of this
geometric flattening, $F$, seen in the power spectrum
$P^S(k_{\parallel},{\mathbf k}_{\perp})$, we can therefore calculate the
true value of $\Omega_{\Lambda}$.

Unfortunately the problem is complicated by the fact that redshift-space distortions are also caused by peculiar velocities. The main cause of redshift-space distortions on the large linear scales probed by the
QSO power spectrum are coherent peculiar velocities due to the infall of galaxies into overdense regions. This anisotropy takes a very simple form in redshift-space, depending only on the density and bias parameters via the combination $\beta \approx \Omega_{\rm m}^{0.6}/b$:
\begin{equation}\label{betaeqn}
P^S(k_{\parallel},{\mathbf k}_{\perp}) =
P^R(k)\left[1+\beta\mu^2\right]^2
\end{equation}
(Kaiser 1987) where $P^S$, and $P^R$ refer to the redshift-space and
real-space power spectra respectively. 

Whilst we expect the parameters
$b$, $\Omega_{\rm m}$, and hence $\beta$ to vary as a function of
redshift, we determine an average $P^S(k_{\parallel},{\mathbf
  k}_{\perp})$ over the redshift range $0.3<z<2.2$ from the 2QZ data, and hence only
fit a single value of $\beta_q(z\sim1.4)$ to the observed
distortion.  We consider the effect of allowing
$\beta_q(z)$ to vary on the model, to test whether such variations could introduce
systematic uncertainties, and whether the
value of $\beta$ obtained represents a true average value for $\beta_q$
over the range $0.3<z<2.2$. Combining the redshift-space
distortion power spectrum models generated using  a varying $\beta_q(z)$, weighted using the QSO $n(z)$, we can
determine an average model $P^S(k_{\parallel},{\mathbf
  k}_{\perp})$. We assume a
simple bias model, designed to match the observations of Outram et
al. (2003), where the QSO clustering amplitude increases
linearly with redshift by 30 per cent from $z=0.3$ to
$z=2.2$ (assuming $\Lambda$). In this evolving bias model, the
best-fitting value of $\beta_q(z)$ increases with
redshift from
$\beta_q(0.3)=0.45$, reaching a maximum of $\beta_q(0.75)=0.51$,
before decreasing to $\beta_q(1.4)=0.45$ and $\beta_q(2.2)=0.34$. The value averaged over the QSO
$n(z)$ is $\bar{\beta_q}=0.44$, in almost perfect agreement with the 
 simple model with a single value
of $\beta_q=0.45$, and the value of  $\Omega_{\Lambda}$ obtained was unchanged, indicating that any systematics introduced by
variations in $\beta$ with redshift are
much smaller than the errors. 

To truly constrain $\beta$ as a function
of redshift, using redshift-space distortions, we would have to measure
the QSO power spectrum in redshift bins. Unfortunately, due to the
smaller sample sizes and volume probed for each bin, the errors in such
an analysis would be considerably larger than any expected change in
$\beta$ with the size of
the current dataset, and so no useful constraint would be obtained. Therefore we have not included such an analysis in this paper.

Anisotropy in the power-spectrum on small scales is dominated by the galaxy velocity dispersions in virialized clusters. This is modelled by introducing a damping term. The line of sight pairwise velocity ($\sigma_p$)\footnote{In power spectra $\sigma_p$ is implicitly divided by $H_0$ and quoted in units $h^{-1}\;$Mpc. ($H_0=100h\;$km$\;$s$^{-1}\;$Mpc$^{-1}$) } distribution is modelled using  a Lorentzian factor in redshift-space:
\begin{equation}
D\left[k\mu\sigma_p\right]=\frac{1}{1+\frac{1}{2}\left(k\mu\sigma_p\right)^2}
\end{equation}
Combining these effects leads to the final model:
\begin{eqnarray}\label{model}
P^S(k_{\parallel},{\mathbf k}_{\perp}) &= &
\frac{P^R(k)}{f_{\perp}^{3+n}F}  
\left[1+\mu^2\left(\frac{1}{F^2}-1\right)\right]^{\frac{n-4}{2}}\nonumber\\
&\times& \left[1+\mu^2\left(\frac{\beta+1}{F^2}-1\right)\right]^2
D\left[k\mu\sigma_p^{\prime}\right]
\end{eqnarray}
(Ballinger et al. 1996) where $\sigma_p^{\prime}=\sigma_p/f_{\parallel}$. 

\section{Mass Clustering Evolution}\label{secmass}
Following Outram et al. (2001), we also consider a second constraint on
$\Lambda$ and $\beta$, based on the
evolution of mass clustering from the average redshift of the 2QZ to
 the present day. 

First we need an estimate of the $z\sim0$ mass clustering
amplitude. For this we could adopt local measurements of $\sigma_8$
(e.g. Eke et al 1996; Hoekstra et al 2002), however, to limit
systematics, we choose to determine the local mass clustering amplitude
using the reverse of the method applied at $z\sim1.4$ using 2QZ. Therefore we will use the recent observations of the 2dF Galaxy Redshift Survey (2dFGRS).

Hawkins et al. (2003) used the 2dFGRS catalogue to determine the real-space galaxy correlation
function at an effective redshift of $z=0.15$, finding it to be well-described by a
power-law $(r/r_0)^{-\gamma_r}$ out to scales of $r\sim 20 h^{-1}$Mpc, where $r_0=5.05 h^{-1}$Mpc and $\gamma_r=1.67$.
Using the same sample of galaxies, Hawkins et al.  produced a measurment of $\beta_{\rm
  g}(z\sim0) = 0.49\pm0.09$, using redshift-space
distortions in the two-point galaxy correlation function. The value of $\beta_{\rm
  g}$ was measured on scales $8<s<30 h^{-1}$Mpc.

For each
cosmology (again we only consider flat cosmologies) the value of the
galaxy-mass bias can be found from $\beta_{\rm g}$, which in turn gives
the amplitude of the mass correlation function from the measured galaxy correlation
function. The evolution in the mass correlation function from $z=0.15$
to $z=1.4$ can then be
calculated for this cosmology using linear theory, and hence by comparing the $z=1.4$
mass correlation function to the 2QZ correlation function, an
estimate of the QSO bias factor, and therefore $\beta_q(z\sim1.4)$ can be obtained,
as a function of cosmology. As in the previous section, this method  assumes that a single value of
$\beta_q$ applies at the average redshift of the QSO survey, which, at
least in the case of the evolving bias model discussed in Section~3, is
a good approximation (we obtain a best fit of $\beta_q(z=1.4)=0.45$ in the evolving
bias model, assuming $\Lambda$, almost identical to the  average value of $\bar{\beta_q}=0.44$).

\section{Results}

\subsection{Fitting the Redshift-space Distortions}
Here we fit the redshift-space distortions model discussed
in Section~\ref{secmod} to the QSO power spectrum,
$P^S(k_{\parallel},\mathbf{k}_{\perp})$, described in
Section~\ref{secpk}. There are several free parameters in the model
(Equation~\ref{model}); $F, \beta, \sigma_p$, and $P^R(k)$, the
underlying real-space power spectrum. To reduce the uncertainties on
the parameters of interest, $\beta$ and $F$, we make simple assumptions about the other parameters.
There is an uncertainty in determining QSO redshifts from low S/N
spectra of $\delta z \sim 0.0038$ (Croom et al. 2003). This introduces
an apparent velocity dispersion of $\sigma_p \sim
670\;$km$\;$s$^{-1}$.  By adding this in quadrature to the intrinsic
small-scale velocity dispersion in the QSO population, we therefore
expect an observed velocity dispersion of $\sigma_p \sim 670 -
840\;$km$\;$s$^{-1}$ (assuming an intrinsic velocity dispersion of
$\sigma_p \sim 0 - 500\;$km$\;$s$^{-1}$).  $\sigma_p$ is not well constrained in this analysis, due to the large
scales probed and so we have allowed $\sigma_p$ to vary, within the
range $670 < \sigma_p < 840\;$km$\;$s$^{-1}$, choosing the
value that maximised the likelihood of the fit. Whilst $\sigma_p$ has relatively little effect on the power spectrum at
large scales, any difference in the true value from our assumed range would lead to a small systematic
shift in the resulting best fit values of $\Omega_{\Lambda}$ and $\beta$.
This could be due, for example,  to an underestimate of
the QSO redshift determination error caused by the intrinsic
variation in line centroids between QSOs.

We reconstruct a real-space power spectrum self-consistently from the
measured QSO redshift-space power spectrum (Outram et al. 2003).  To
remove noise we
adopt the best-fitting model redshift-space power spectra given in Table 4 of Outram
et al.. The real-space power spectrum is then obtained by
inverting the above redshift distortion equations in each cosmology for
an average value of $\mu$. Finally, we assume a flat cosmology,
consistent with the recent WMAP CMB results (Spergel et al. 2003). We choose to fit the variable $\Omega_{\Lambda}$, and fix $\Omega_{\rm m}=1-\Omega_{\Lambda}$.

We fit the model to the 2QZ data by performing a maximum likelihood
analysis, allowing the parameters $\beta$, and $\Omega_{\Lambda}$ to vary
freely and determining the likelihood of
each model by calculating the $\chi^2$ value of each fit.
Only those wavenumbers with $k_{\parallel}>0.02, |{\mathbf
  k}_{\perp}|>0.02$, and $k<0.2$ (assuming $\Lambda$), or
$k_{\parallel}>0.03, |{\mathbf k}_{\perp}|>0.03$, and $k<0.3$ (assuming
EdS) are used in the fit. The former constraints are applied to remove
the effects of the window function, and the latter because the FFT is
unreliable at smaller scales. This constraint also prevents excessive
non-linearity, where the model breaks down.

\begin{figure}
\vspace{-0.5cm}
\centerline{\hbox{\psfig{figure=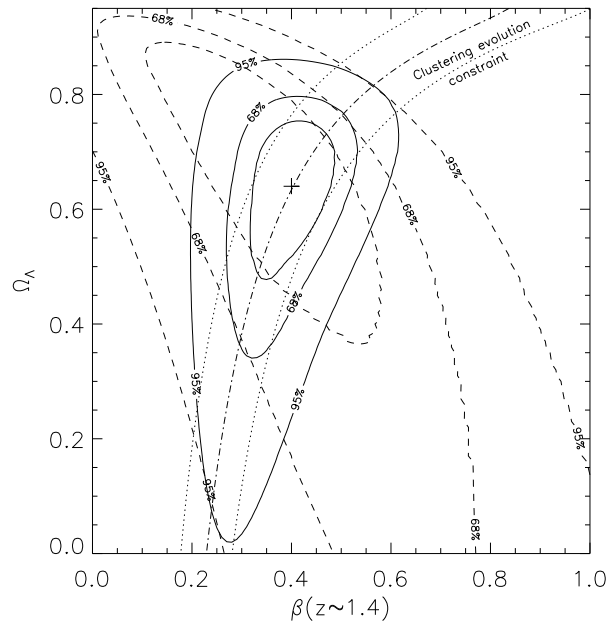,width=9cm}}}
\vspace{-0.3cm}
\caption{Likelihood contours are
  plotted  in the
  $\Omega_{\Lambda}$ -- $\beta$ plane for $\chi^2$ values corresponding to a one-parameter confidence of 68 per cent, and
  two-parameter confidence of 68 and 95 per cent (dashed contours),
  calibrated using Monte Carlo simulations, for fits to
  $P^S(k_{\parallel},\mathbf{k}_{\perp})$ determined from the 2QZ
  catalogue assuming an EdS  $r(z)$.
Overlaid are the best-fit (dot-dash) and 1-$\sigma$ (dot) values of  $\beta$ determined using the mass clustering evolution method.
Significance contours given by joint consideration of the two
constraints are also plotted for a one-parameter confidence of 68 per
cent, and two-parameter confidence of 68 and 95 per cent (solid
contours). The best fit model obtained (marked with a $+$) has $\beta = 0.40$ and $\Omega_{\Lambda} = 0.64$.
}
\label{figfitom}
\end{figure}

\begin{figure}
\vspace{-0.5cm}
\centerline{\hbox{\psfig{figure=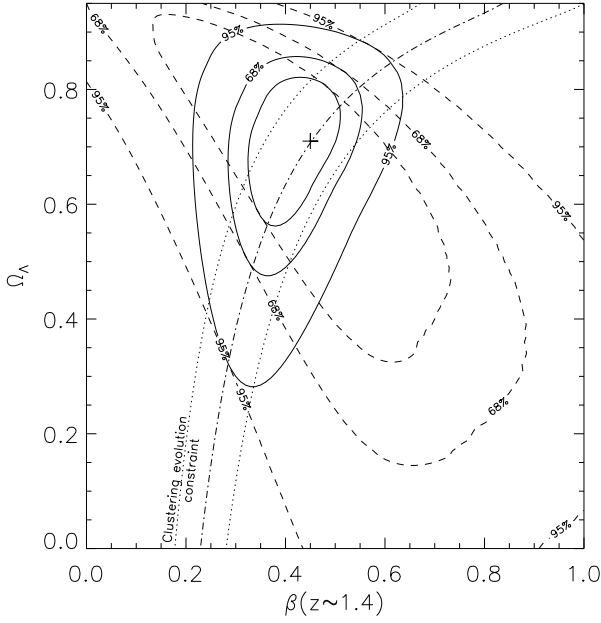,width=9cm}}}
\vspace{-0.3cm}
\caption{Likelihood contours are
  plotted in the
  $\Omega_{\Lambda}$ -- $\beta$ plane for $\chi^2$ values corresponding to a one-parameter confidence of 68 per cent, and
  two-parameter confidence of 68 and 95 per cent (dashed contours),
  calibrated using Monte Carlo simulations, for fits to
  $P^S(k_{\parallel},\mathbf{k}_{\perp})$ determined from the 2QZ
  catalogue assuming a $\Lambda$ $r(z)$.
Overlaid are the best-fit (dot-dash) and 1-$\sigma$ (dot) values of  $\beta$ determined using the mass clustering evolution method.
Significance contours given by joint consideration of the two
constraints are also plotted for a one-parameter confidence of 68 per
cent, and two-parameter confidence of 68 and 95 per cent (solid
contours). The best fit model obtained (marked with a $+$) has $\beta = 0.45$ and $\Omega_{\Lambda} = 0.71$.
}
\label{figfitlam}
\end{figure}

Figures~\ref{figfitom} and \ref{figfitlam} show likelihood contours in
the $\Omega_{\Lambda}$ -- $\beta$ plane, assuming an EdS and $\Lambda$
cosmology $r(z)$ respectively. Nominally the best fit values
obtained from the redshift-space distortions analysis are
$\beta=0.30^{+0.28}_{-0.19}$, and $\Omega_{\Lambda}=0.76^{+0.13}_{-0.39}$, with $\chi^2=61.5$ over 57
degrees of freedom assuming EdS, or $\beta=0.43^{+0.29}_{-0.30}$, and
$\Omega_{\Lambda}=0.73^{+0.20}_{-0.40}$, with $\chi^2=54.1$ over 61
degrees of freedom assuming $\Lambda$. In both cases, the absolute $\chi^2$
values indicate that the best-fitting models provide an adequate fit to
the power spectrum. However, when comparing the absolute $\chi^2$
values from different power spectrum realisations, we caution that
there is considerable ($\sim 10$ per cent) noise in the $\chi^2$ values
obtained, which manifests itself even  with the same
cosmology and selection function, when only subtle
changes to the FFT parameters or binning are made.
As Figures~\ref{figfitom} and~\ref{figfitlam} show, there is a large
degeneracy between the fitted values of the two parameters, due to the
similarity in the shape of the redshift-space and geometric distortions
(Ballinger et al. 1996).


\subsection{Testing the likelihood contours}
\begin{figure}
\vspace{-0.5cm}
\centerline{\hbox{\psfig{figure=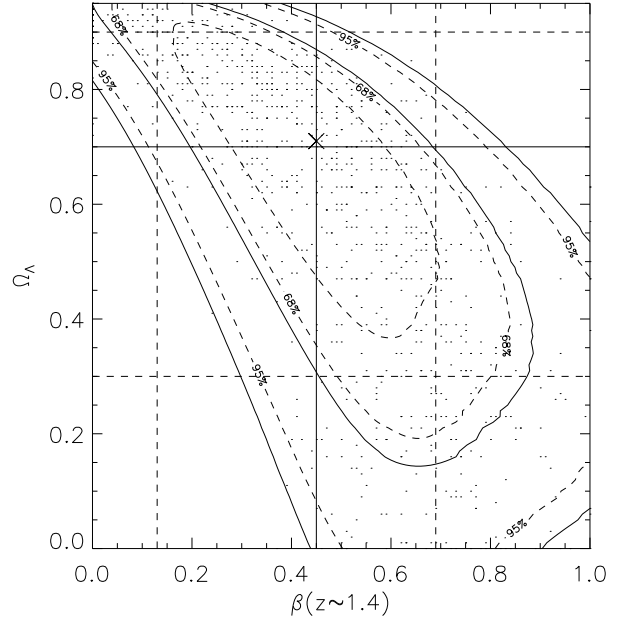,width=9cm}}}
\vspace{-0.3cm}
\caption{A comparison between the two-parameter confidence of 68 and 95 per cent likelihood contours (dashed), derived using the
  $\chi^2$ statistic from a fit to  $P^S(k_{\parallel},\mathbf{k}_{\perp})$ measured from the 2QZ
  catalogue assuming a $\Lambda$ $r(z)$, and solid contours derived
  from 1000 Monte Carlo simulations of
  $P^S(k_{\parallel},\mathbf{k}_{\perp})$ drawn from the best fit model
  (marked with a $+$), assuming the fractional errors on the power
  spectrum calculated  from the dispersion in the measurements from 
 $5^{\circ} \times 10^{\circ}$ regions. The points show the best fit
 values for $\Omega_{\Lambda}$ and
  $\beta$ obtained from each realization of the power spectrum. The
  straight solid and dashed lines show the median  and 16/84 percentile
  values for $\Omega_{\Lambda}$ and
  $\beta$ from the Monte Carlo simulations, which can be compared to
  the inner dashed contour showing the one-parameter 68 per cent
  confidence likelihood contour.
}
\label{errlam}
\end{figure}

To test and calibrate the likelihood contours and hence the parameter uncertainties
derived from the $\chi^2$ model fitting to the 2QZ
$P^S(k_{\parallel},\mathbf{k}_{\perp})$, Monte Carlo simulations were
performed. 1000 realizations of the $\Lambda$ $r(z)$ 2QZ power spectrum
were drawn from the best fitting model ($\beta=0.45$, $\Omega_{\Lambda}=0.71$), assuming a fractional uncertainty on
each data point equal to that estimated from the true power spectum
(see Section~\ref{errsec}). The redshift-space distortions model was
fitted to each realization of the QSO power spectrum, as discussed
above. The best fit values for $\Omega_{\Lambda}$ and
  $\beta$ obtained from each realization of the power spectrum are
  plotted in Fig.~\ref{errlam}, together with the median and 16/84 percentile
  values of each parameter. The best fit values
obtained from the Monte Carlo redshift-space distortions analysis are
$\beta=0.45^{+0.24}_{-0.32}$, and $\Omega_{\Lambda}=0.70^{+0.20}_{-0.40}$, in good agreement with the input model
to the simulations ($\beta=0.45$, $\Omega_{\Lambda}=0.71$), however, the size of the uncertainties estimated via the Monte
Carlo analysis are slightly larger than those estimated via the $\chi^2$
likelihood analysis. The $\chi^2$ contours containing 68 and 95 per cent of the simulated measured parameters were also calculated
  from the simulations, and
  are compared to the equivalent $\chi^2$ likelihood contours in
  Fig.~\ref{errlam}. Again, the Monte Carlo contours appear fairly
  close, but slightly larger than the
  likelihood contours.  The Monte Carlo simulations indicate that the
  likelihood contours underestimate the $\Delta \chi^2$ values
  corresponding to a given confidence level by approximately 20 per
  cent, and so for this analysis we have corrected the contour levels
  by this factor.

\subsection{The Mass Clustering Evolution Constraint}

We now combine the degenerate redshift-space distortion constraint
obtained from the likelihood analysis above with the constraint obtained from
consideration of the evolution of mass clustering, described in Section~\ref{secmass}.
The value of $\beta_q(z\sim1.4)$ and the one-sigma errors from this constraint
are plotted on
figures~\ref{figfitom} and \ref{figfitlam}. Although totally degenerate
with the value of $\Omega_{\Lambda}$, this provides a different, and
almost orthogonal constraint to that from the redshift-space
distortions. This was derived using significantly smaller scales than
the $P^S(k_{\parallel},\mathbf{k}_{\perp})$ analysis, and so we can
treat the results as independent and hence combine the likelihoods,
yielding a much stronger fit.

Assuming an EdS $r(z)$, the joint best fit values obtained are  $\Omega_{\Lambda}=0.64^{+0.11}_{-0.16}$ and
$\beta_q(z\sim1.4)=0.40^{+0.09}_{-0.09}$. The best fitting flat $\Omega_{\Lambda} = 0$ model with
  $\beta = 0.27$ (shown in Figure~\ref{figdataom}) can be excluded at over 95 per cent confidence. 
The joint best fit model to the $\Lambda$ $r(z)$ 2QZ data has
$\Omega_{\Lambda}=0.71^{+0.09}_{-0.17}$ and
$\beta_q(z\sim1.4)=0.45^{+0.09}_{-0.11}$. Likelihood contours in
the $\Omega_{\Lambda}$ -- $\beta$ plane for the two fits are
  plotted for a  one-parameter confidence of 68 per cent, and
  two-parameter confidence of 68 and 95 per cent in
  figures~\ref{figfitom} and \ref{figfitlam}. 

By combining this constraint with that derived from the redshift-space
distortions, we are comparing values of $\beta$ that were measured using different estimators and on
different scales;  $r\sim 20 h^{-1}$Mpc for the mass clustering
evolution method described in this section, compared to $r\sim 100
h^{-1}$Mpc for the redshift-space distortions in the power spectrum. 
Hence we are implicitly assuming that bias is scale
independent on these scales. Whilst this is likely to be true (e.g. Verde et
al. 2002), a non-linear bias on these scales would
introduce a systematic error in our results.
To reduce any possible systematic effects further the value
of $\beta_g$ could be measured using the same method (Outram, Hoyle \&
Shanks 2001), however this analysis has yet to be done using the large
2dF galaxy sample. Although at a much lower redshift, there will also still
be a small cosmological dependence in the determined value of $\beta_g$
which should be taken into account.

Instead of using the correlation
function amplitude to trace the evolution in clustering, the amplitude of the spherically averaged QSO power
spectrum (Outram et al. 2003) could be compared to that of local
galaxies (Percival et al. 2001). The different effects on the respective power
spectra of small-scale velocity dispersions, redshift uncertainties and, most importantly, the two
survey window functions have to be taken into account in the
comparison. The published 2dFGRS power spectrum was measured
assuming the $\Lambda$ cosmology $r(z)$, so comparing its amplitude to that of
the $\Lambda$ $r(z)$ QSO  power
spectrum, and also assuming, as before, $\beta_{\rm
  g}(z\sim0) = 0.49\pm0.09$ we would predict $\beta_q=0.43\pm0.09$,
in good agreement with the prediction of $\beta_q=0.45\pm0.09$ obtained
using the correlation function analysis. 

Alternatively, we could choose to constrain the local mass clustering
 amplitude using the relation $\sigma_8=
(0.52+/-0.04)\Omega_m^{- 0.52+0.13\Omega_m}$
 (Eke et al. 1996), determined from the evolution of rich galaxy
 clusters, instead of using the 2dFGRS observations. With this approach we again obtain very similar results, with $\Omega_{\Lambda}=0.75^{+0.08}_{-0.17}$ and
$\beta_q(z\sim1.4)=0.41^{+0.09}_{-0.09}$.

\section{Comparison with the Hubble Volume simulation}

\begin{figure}
\vspace{-0.5cm}
\centerline{\hbox{\psfig{figure=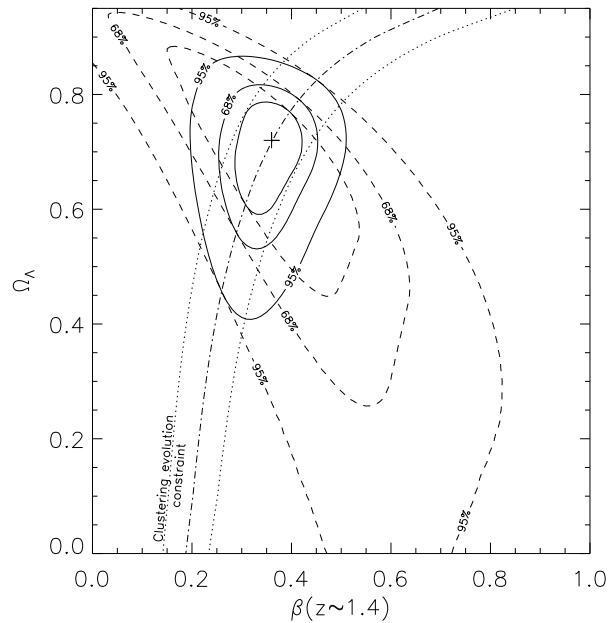,width=9cm}}}
\vspace{-0.3cm}
\caption{Contours are plotted in the $\Omega_{\Lambda}$ -- $\beta$ plane for a one-parameter confidence of 68 per cent, and two-parameter confidence of 68 and 95 per cent (dashed contours) for fits to
  $P^S(k_{\parallel},\mathbf{k}_{\perp})$ determined from the 25k
  Hubble Volume mock QSO
  catalogue assuming a $\Lambda$ $r(z)$. 
Overlaid are the best-fit (dot-dash) and 1-$\sigma$ (dot) values of  $\beta$ determined using the mass clustering evolution method.
Significance contours given by joint consideration of the two
constraints are also plotted for a one-parameter confidence of 68 per
cent, and two-parameter confidence of 68 and 95 per cent (solid
contours). The best fit model obtained (marked with a $+$) has $\beta = 0.36$ and $\Omega_{\Lambda} = 0.72$.
}
\label{figfithv}
\end{figure}

Outram et al. (2001) produced an analysis of
redshift-space distortions in the power spectrum,
$P^S(k_{\parallel},\mathbf{k}_{\perp})$, of a
simulation of the 2QZ, created using the Virgo
Consortium's huge Hubble Volume N-body $\Lambda$CDM simulation
(Frenk et al. 2000, Evrard et  al. 2002).   
The parameters of the simulation are $\Omega_{\rm b}$=0.04,
$\Omega_{\rm CDM}$=0.26, $\Omega_{\Lambda}$=0.7, $H_{\circ}$=70 km
s$^{-1}$Mpc$^{-1}$ and the normalisation, $\sigma_8$, is 0.9. 
The simulation can be used to produce three light-cone 75$\times$5
degree declination strips extending to $z \sim $4. A simple biasing prescription
and sparse sampling is
used to give the mass particles a similar clustering pattern and
selection function as the final 2QZ survey.

As we have refined the method of analysis for this paper, for a full
comparison with the results presented here we repeat the
analysis of the Hubble Volume simulation,
using exactly the same method  as for the final 2QZ catalogue. Two
$5\times75\deg^2$ declination strips, each containing 12500 mock QSOs,
were used in the analysis; see Outram et al. for further details.
To match the 2QZ analysis, the mock QSO redshifts were degraded to the
noise level from 2QZ ($\delta z \sim 0.0038$, corresponding to an apparent velocity dispersion of $\sigma_p \sim 670$km$\;$s$^{-1}$). As before, $\sigma_p$ was allowed to vary, within the
range $670 < \sigma_p < 840\;$km$\;$s$^{-1}$, choosing the
value that maximised the likelihood of the fit.

The results are shown in Fig.~\ref{figfithv}. The joint best
fit model to the $\Lambda$ $r(z)$ mock QSO data has
$\Omega_{\Lambda}=0.72^{+0.07}_{-0.13}$ and
$\beta_q(z\sim1.4)=0.36^{+0.07}_{-0.07}$. These results are consistent
both with the input values for the simulation, $\beta_q = 0.35$ and
$\Omega_{\Lambda} = 0.7$, and the best fitting values obtained by Outram
et al. (2001). The uncertainty in the estimates of the two parameters,
however,  are considerably smaller than those quoted by Outram et
al.. This is because they incorrectly calculated the one-parameter
uncertainties for $\beta$ and $\Omega_{\Lambda}$ from the two-parameter $\chi^2$
contours, significantly overestimating the true uncertainty. The two-parameter 68 and 95 per cent confidence levels, however,
are very similar to those derived by Outram et al.. Using the Hubble
Volume simulation, and the uncertainty estimates derived from the one-parameter 68
per cent likelihood contour, we would instead predict that this method should
constrain $\beta$ to approximately $\pm0.07$, and $\Omega_{\Lambda}$ to
$\pm0.1$ using the final 2QZ catalogue. 

In the 2QZ catalogue there remains a small level of incompleteness that
was not considered in the simulation, and therefore there are approximately 2000
fewer QSOs. The uncertainties inherent in constructing a completeness
mask to correct for this in the 2QZ will also add a small amount of
noise that was not considered in the simulation, although Outram et
al. (2003) demonstrated that any potential systematics due to
this are significantly smaller than the
statistical errors in the power spectrum determination.
Hence, the uncertainties quoted above are slightly
optimistic. However, the robust way that the Hubble Volume analysis returned
almost exactly the true input parameters, coupled with remarkable similarity
between both the results and the uncertainties
from the simulation, shown in Fig.~\ref{figfithv}, and the data, shown
in Fig.~\ref{figfitlam}, gives us great confidence in the results we
present here from the 2QZ catalogue.

\section{Discussion and Conclusions}

In this paper we have reported on measurements of the cosmological constant,
$\Lambda$, and the redshift space distortion parameter
$\beta=\Omega_m^{0.6}/b$, based on an analysis of the QSO power
spectrum parallel and perpendicular to the observer's line of sight, 
$P^S(k_{\parallel},\mathbf{k}_{\perp})$, from the final catalogue of
the 2dF QSO Redshift Survey.
We have derived a joint $\Lambda - \beta$ constraint  from the geometric and
redshift-space distortions in the power spectrum. By combining this
result with a second constraint based on mass clustering evolution, obtained by comparing the clustering amplitude of 2QZ QSOs at $z\sim1.4$ with that of 2dFGRS galaxies at $z\sim0$ (Hawkins et al. 2003), we
have broken this degeneracy, obtaining strong constraints on both parameters.

Assuming a flat ($\Omega_m+\Omega_{\Lambda}=1$) cosmology and a $\Lambda$ cosmology $r(z)$ we find
best fit values of 
$\Omega_{\Lambda}=0.71^{+0.09}_{-0.17}$ and
$\beta_q(z\sim1.4)=0.45^{+0.09}_{-0.11}$. Assuming instead an EdS cosmology
$r(z)$ we find that the best fit model obtained, with $\Omega_{\Lambda}=0.64^{+0.11}_{-0.16}$ and
$\beta_q(z\sim1.4)=0.40^{+0.09}_{-0.09}$, is consistent with the $\Lambda$ $r(z)$
results, and again strongly favours a
  $\Lambda$ dominated cosmology. Indeed, the EdS cosmology
$r(z)$ result is inconsistent with a flat
$\Omega_{\Lambda}=0$ cosmology at over 95 per cent confidence. 

We determine the 2QZ power spectrum in a single adopted cosmology.
Following Alcock \& Paczy\`{n}ski (1979), Ballinger et al. (1996) and
Outram et al. (2001),
we fit the size of the geometric distortions in the power spectrum that arise if this
cosmology is incorrect, hence deriving a constraint on the true value
of $\Omega_{\Lambda}$. An alternative approach would be to measure the
2QZ power spectrum in each of the cosmologies considered in this paper
(with $0\le \Omega_{\Lambda}\le1$),
rather than just an adopted cosmology, then measure the
goodness-of-fit of the redshift-space distortions model (without
geometric distortions) in that cosmology, determining which cosmology
provides the best overall fit. Obviously, this
would be expected to occur when there were no geometric distortions in
the power spectrum. A comparison of the $\chi^2$ values obtained for
the best fits to the $\Lambda$ and EdS cosmology 2QZ power spectra
suggests that this method might provide an even tighter constraint on
$\Omega_{\Lambda}$. However, there is considerable ($\sim 10$ per cent)
noise in the absolute $\chi^2$ value of the best fit to any power
spectrum realisation (which manifests itself even when only subtle
changes to the FFT parameters or binning are made, let alone changes in
the adopted cosmology). Whilst this has little effect on the best
fitting $\beta$ and $\Omega_{\Lambda}$ parameters in a given
realisation (much smaller than the quoted uncertainties), it does
inhibit the comparison of $\chi^2$ fits from different realisations.
This approach would no doubt also confirm that $\Lambda$ dominated
cosmologies are strongly favoured by the 2QZ power spectrum, however, a much noisier picture would therefore emerge.

 Outram et al. (2003) presented an analysis of the 
  spherically-averaged 2QZ power spectrum.  They found very good agreement
  between the QSO power spectrum and the Hubble Volume $\Lambda$CDM simulation over the
  whole range of scales considered.
From the shape of the QSO power
  spectrum Outram et al. derived a low
  value of $\Omega_mh=0.19\pm0.05$ (where $h$ is the Hubble parameter) that is hard to reconcile with a standard ($\Omega_{\rm m}=1$) CDM
  model.
 Their results are entirely consistent with the analysis described in this paper, of redshift-space distortions in the same QSO power spectrum, 
  and taken together they provide strong evidence in favour of
  the $\Lambda$CDM scenario.

\section*{Acknowledgements} 2QZ was based on observations made with the
Anglo-Australian Telescope and the UK Schmidt Telescope, and we would
like to thank our colleagues on the 2dF Galaxy Redshift Survey team and
all the staff at the AAT that have helped to make this survey
possible. We would like to thank Adam Myers for useful discussions and
the anonymous referee for helpful comments.
This work was partially supported by the SISCO European Community Research
and Training Network. 
PJO would like to acknowledge the support of a PPARC Fellowship.

\end{document}